\documentclass[12pt,preprint]{aastex}
\begin{document}

\title{Mass Distribution of Spiral Galaxies in a Thin Disk Model with Velocity Curve Extrapolation}
\shorttitle{thin disk model with velocity extrapolation}
\author{Valentin Kostov}
\affil{University of Chicago}
\affil{Department of Physics, 5640 S. Ellis Ave., Chicago, IL 60637}
\email{valentin@uchicago.edu}
\begin{abstract}

We model a spiral galaxy by a thin axially symmetric disk that includes both visible and dark matter. The surface mass density of the disk is calculated directly from the rotational velocity curve without extra assumptions. We simplify the standard application of the model. Since most velocity curves are known out to some radius, $r_{max}$, we extrapolate them by attaching a Keplerian tail. The numerical procedure and the extrapolation are tested with a known toy mass density and shown to reconstruct it with a good precision if $r_{max}$ includes a sufficient part of the velocity curve. Mass density curves are calculated for Milky Way and NGC 3198. We vary the extent of the flat part of the velocity curves from $30\;kpc$ to $200\;kpc$ and show that does not affect appreciably the calculated mass density inside $r_{max}=30 \; kpc$ . The reconstructed masses for Milky Way are $15 \times 10^{10} \; M_\odot$ inside the visible disk and $23 \times 10^{10} \; M_\odot$ inside $30 \; kpc$. For NGC 3198, the reconstructed mass inside the visible disk is $6.5 \times 10^{10} \; M_\odot$ and $11 \times 10^{10} \; M_\odot$  inside $30 \; kpc$. The total galactic masses are roughly proportional to the extent of the flat part of the velocity curves which is currently unknown. The high  light-to-mass ratios obtained for the visible disks of the galaxies - $11 \; M_\odot/L_{V\odot}$ for Milky way and $9.3 \; M_\odot/L_{V\odot}$ for NGC 3198 - suggest presence of dark matter. The method  is also applied to NGC 3031 - a spiral galaxy with a declining velocity curve in which case it is able to reconstruct both the mass density curve and the total mass ($14 \times 10^{10} \; M_\odot$).

\end{abstract}

\section{INTRODUCTION}
The flat rotational velocity curves of spiral galaxies remain a central problem in astrophysics. Exotic attempts for explanation were made in terms of Modified Newtonian Dynamics \citep{mond} or even incorporating general relativity effects \citep{cop}. The most popular approach remains a three component galaxy model consisting of a bulge, a thin disk, and a spherical dark matter halo \citep{kent2}. Usually, the mass-to-light ratios of the bulge and the disk are taken constant and a the halo mass density is assumed to vary as $1/[1+(r/a)^2]$ since that produces a flat velocity curve by itself. The several parameters of the model are then fit to the observed velocity profile. Unfortunately, such fits do not produce unique values of the parameters and frequently some of them have to be chosen before the fit \citep{textbook}. A model with fewer a priori assumptions is desirable.

In this paper, a spiral galaxy is represented by an axially symmetric thin disk of surface mass density $\sigma(r)$ which includes all types of matter. It is very natural to assume that the dark matter component(if any) resides in the galaxy disk just like the visible matter since both are supposed to have the same gravitational properties. The advantage of that model is that $\sigma(r)$ can be calculated directly from the velocity curve $v(r)$ without invoking extra assumptions. The necessary formulas were developed by \citet{toom}. Since the velocity curves are known out to a radius $r_{max}$, they have to be extrapolated by a Keplerian decline \citep{nord}. The thin disk model is applicable in cases where the galaxy radius is much bigger than its thickness and the size of the bulge. It ignores deviations from axial symmetry due to the spiral arms or possible bars. Newtonian gravity law is used throughout assuming the general relativity effects are negligible which is the prevalent opinion in the field. The model does not address the question why the mass is distributed in the way calculated.

In the current paper, we apply the model in a way that in our opinion is simpler than the standard approach. Using a toy mass density, we test numerically how the accuracy of the reconstructed mass density  depends on the point, $r_{max}$, after which the velocity profile is extrapolated. At the end, the density curves of several galaxies are calculated using reasonable guesses of the extent of the flat part of the velocity curves. Luckily, the density curves and the masses inside the current experimental $r_{max}$ are not very sensitive to that extent. On the other hand, only a range of the total masses of the galaxies can be given, since they are roughly proportional to the unknown extent of the velocity curve flatness.

\section{RELATION BETWEEN MASS DENSITY AND VELOCITY CURVE FOR A THIN DISK}
Outside the rotating disk, the gravitational potential satisfies the Laplace equation
\begin{equation}
\nabla^2\Phi=0
\end{equation}
whose general solution in cylindrical coordinates for $z>0$ is \citep[p.118]{jak} 
\begin{equation}
\Phi(r, \phi,z) = \sum_{m=0}^{\infty} \int_0^{\infty} dk \;e^{-kz} J_m(kr)[A_m(k)\cos m\phi + B_m(k)\sin m\phi]
\end{equation}
where $J_m$ are the usual Bessel functions of the first kind and $A_m$ and $B_m$ are expansion coefficients.
The solution for $z<0$ is just a mirror image of the above.
An axially symmetric solution is given by setting $m=0$:
\begin{equation}
\Phi(r,z) =\int_0^{\infty} dk\; e^{-kz} J_0(kr)A(k)
\end{equation}
where $A(k)\equiv A_0(k)$ denote the Bessel spectrum of the potential. 
Assuming a finite mass density, the gravitational potential must be continuous on the disk plane, $z=0$, and is given by
\begin{equation}
\Phi(r,z=0) =\int_0^{\infty} dk \;J_0(kr) \;A(k).
\end{equation}
This equation can be inverted
\begin{equation}
A(k) =k \int_0^{\infty} dr\; r \; J_0(kr)\; \Phi(r,z=0)
\end{equation}
using the orthogonality relation \citep[p.118]{jak}
\begin{equation}
\int_0^{\infty} dr\; r J_m(kr)J_m(k'r) = \frac{1}{k}\delta(k-k').
\end{equation}

The rotational velocity profile, $v(r)$, is connected to the potential by 
\begin{equation}
\frac{v^2(r)}{r} = \frac{\partial\Phi(r,0)}{\partial r}= - \int_0^{\infty} dk \; k \;J_1(kr)A(k) 
\end{equation}
where we used $dJ_0(x)/dx = -J_1(x)$.
This can be inverted using equation (6) with m=1 giving
\begin{equation}
A(k) = - \int_0^\infty  dr\; J_1(kr) \; v^2(r)   .
\end{equation}

On the other hand, Gauss theorem connects the surface mass density, $\sigma(r)$, to the potential by
\begin{equation}
\sigma(r) =  \frac{2}{4\pi G} \frac{\partial\Phi(r,z)}{\partial z}\Biggl |_{z\rightarrow 0+}= 
\frac{1}{2\pi G}\int_0^{\infty} dk \; (-k) \; J_0(kr)A(k), 
\end{equation}
where $G$ is the gravitational constant.
The Bessel functions $J_0(kr)$ form a complete set and the mass density can be expanded analogously to the potential
\begin{equation}
\sigma(r) =  \int_0^\infty dk\; J_0(kr) \; \sigma(k)   
\end{equation}
where $\sigma(k)$ is the Bessel spectrum of the density.
Comparing the last two equations gives the relation between the Bessel spectra of the potential and 
the surface mass density
\begin{equation}
\sigma(k) =  -\frac{kA(k)}{2\pi G} .
\end{equation}

The way to compute the mass density for a given velocity profile, $v(r)$, is now clear: 

1. Use equation (8) to compute $A(k)$, the Bessel spectrum of the gravitational potential.

2. Obtain the Bessel spectrum of the mass density, $\sigma(k)$ from equation (11).

3. Reconstruct the surface mass density, $\sigma(r)$, from equation (10).

\section{EXTRAPOLATING THE VELOCITY CURVE}
The velocity curves of galaxies are known from HI spectroscopy out to some radius, $r_{max}$, of about $20-40\; kpc$. They have to be extrapolated to $r=\infty$ before computing equation (8).

The most elementary approach used in \citet{soph} would be to perform the integration in equation (8) up to $r_{max}$ . That is equivalent to setting the rotational speed to zero for bigger radii. Unfortunately, as the calculated $\sigma(r)$ has shown, that produces a ring of negative mass density around $r_{max}$ which is necessary to cancel out the field of the positive mass so that $v(r)$ can drop to zero at that radius. Negative mass density is hard to justify physically. Another possibility is to fit the velocity profile and let the continuation of the fitting function determine the behavior for $r>r_{max}$. That is not acceptable either because the fit continuation usually does not have the necessary Keplerian decline at bigger radii.

If we assume that $r_{max}$ is close to the Keplerian tail of the velocity curve, we can estimate the total mass of the disk, $M_e$, using the point-mass formula 
\begin{equation}
M_e =\frac{r_{max} \; v^2(r_{max})}{G}.
\end{equation}
That estimate is good if all or a big part of the mass is inside $r_{max}$.
In this paper, we extrapolate the velocity curve by attaching a Keplerian tail to it: 
\begin{equation}
v(r) = \sqrt{\frac{G M_e}{r}}, \;\; r>r_{max},
\end{equation}
a similar approach was used by \citet{nord}.

Numerical integration of Bessel functions over a large range like in equation (8) can give severe cancellation errors. That is why all authors \citep{toom, nord, soph} avoided the simple Bessel spectrum formulation presented in the previous section and would rather integrate over the velocity derivative $dv^2(r)/dr$ and use elliptic integrals. That has the inconvenience that the elliptic integrals are divergent at a point and the velocity derivative has to be computed numerically from the velocity curve which is a possible source of errors. The success of our computations shows all these complications are not necessary since we take the integral of equation (8) numerically up to $r=r_{max}$ and analytically after that:
\begin{equation}
\int_{r_{max}}^\infty  dr\; J_1(kr) \; v^2(r)  =  G M_e \int_{r_{max}}^\infty  dr\; \frac{J_1(kr)}{r}  = 
G M_e \cases{ 
1-\frac{k r_{max}}{2} \ F[\frac{1}{2}, \{ \frac{3}{2},2\}, -(\frac{k r_{max}}{2})^2)]\; & $, k \ne 0$ \cr
0 & $, k=0$ \cr}
\end{equation}
Here $F[\cdots]$ denotes the generalized hypergeometric function for which good numerical routines exist. It is defined by the expansion
\begin{equation}
F[a, \{b_1, b_2\}, z]=\sum_{n=0}^\infty \frac{\frac{\Gamma(a+n)}{\Gamma(a)}}{\frac{\Gamma(b_1+n)}{\Gamma(b_1)}\frac{\Gamma(b_2+n)}{\Gamma(b_2)}}
\frac{z^n}{n!},
\end{equation}
where $\Gamma(x)$ denotes the usual gamma function.

\section{CALCULATION PROCEDURE}

The velocity curve is a smooth fit to the experimental points out to $r_{max}$. It is extrapolated according to equations (12) and (13). The total mass of the computed $\sigma(r)$ will be $M_e$ since it is already encoded in the extrapolated Keplerian tail of $v(r)$. The gravitational potential Bessel spectrum, $A(k)$, of  equation (8) is sampled for $k=0 ... k_{max}$ with a step of $\Delta k$. The corresponding mass density spectrum, $\sigma (k)$, is computed from equation (11) and interpolated with splines. The values for $k_{max}$ and $\Delta k$ are chosen so that the computed spectrum is without sudden kinks and stabilizes to zero before $k_{max}$. The spline-fitted $\sigma (k)$ is integrated numerically according to equation (10) returning $\sigma(r)$ at the points of interest.

\section{TESTING THE METHOD WITH A TOY MASS DENSITY}
In this section we work in arbitrary units and the gravitational constant $G=1$.
Our toy surface mass density shown on Fig.1(a) is
\begin{equation}
\sigma(r) =  \cases{ 
5.5+0.5 \; cos(2\pi r) & $, 0<r\le 2$ \cr
\frac{6}{1+(r-2)^2} & $, 2<r\le 4$ \cr
1.2 & $, 4<r\le 6$ \cr
4.8-0.6 \; r& $, 6<r\le 8$ \cr
0 & $, r>8$ \cr
}
\end{equation}

\begin{figure}
\includegraphics[width=7.1cm]{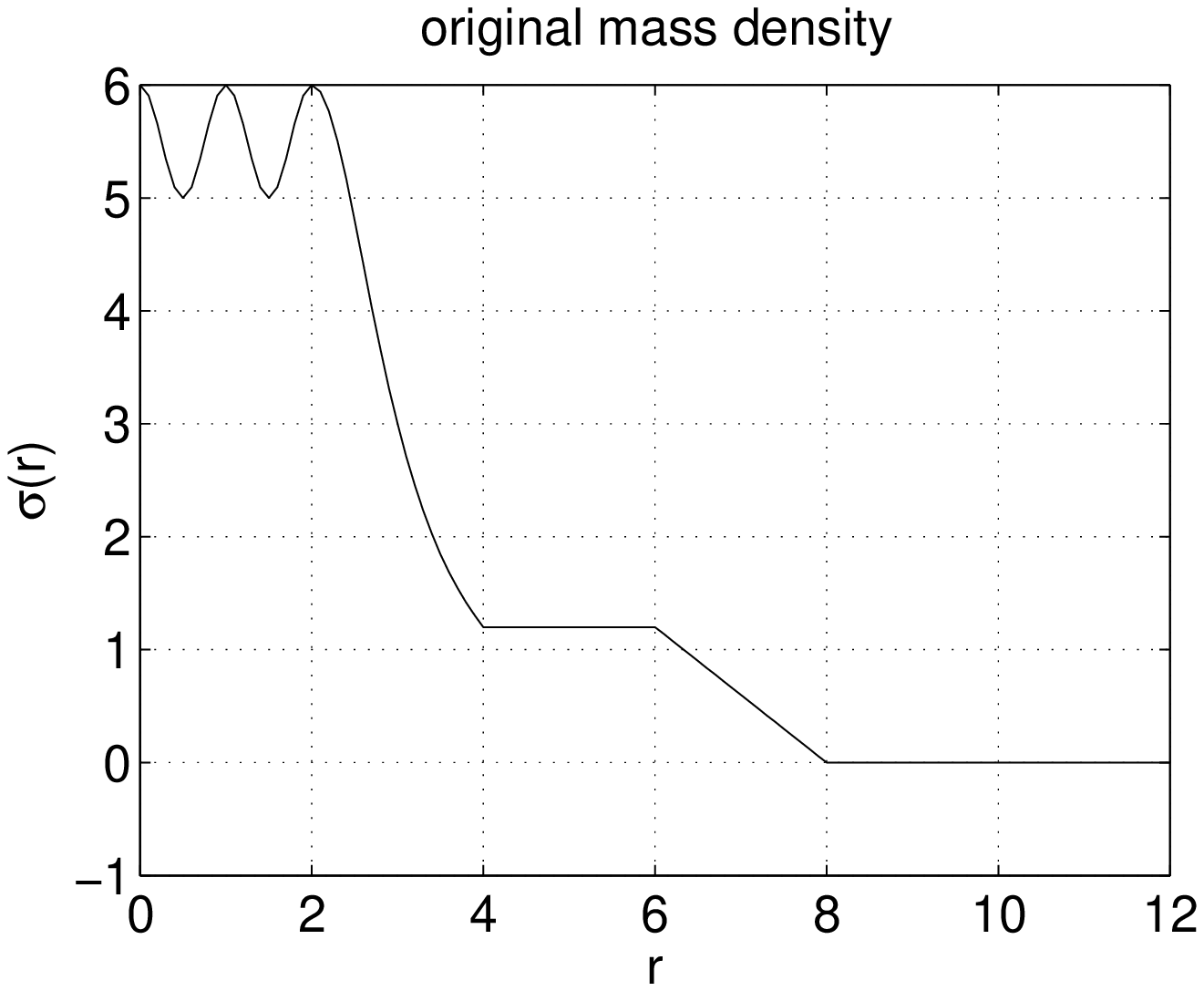}
\includegraphics[width=7.1cm]{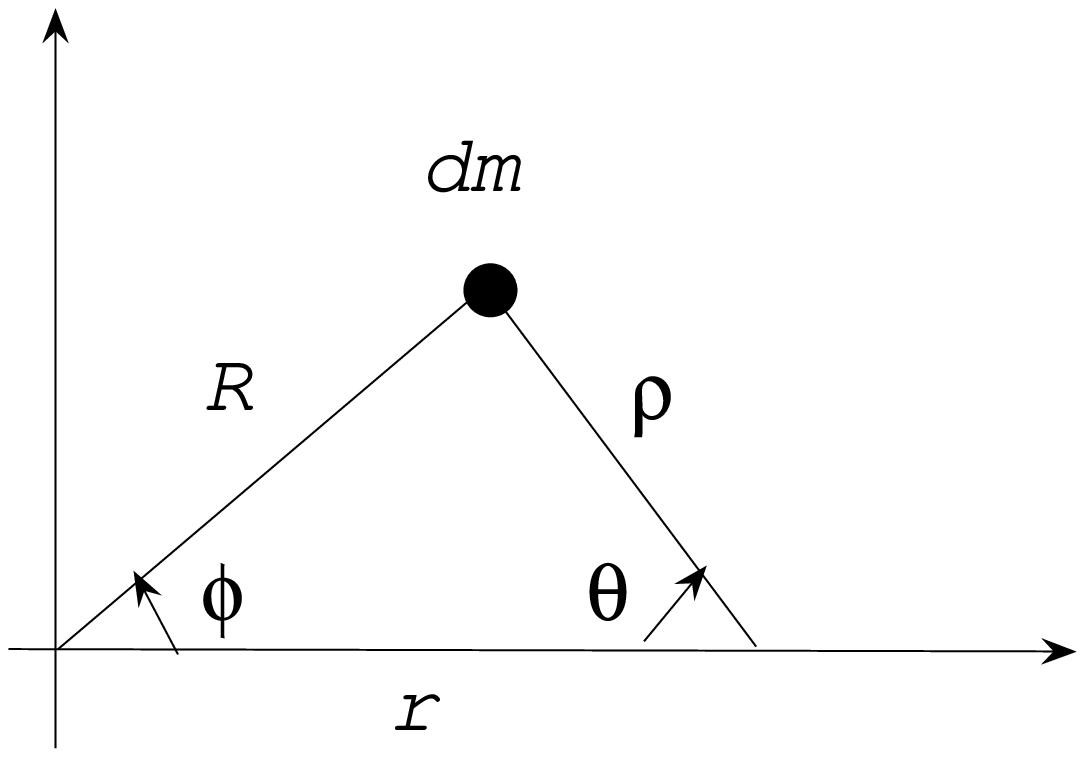}
\caption{(a) original (test) mass density  
(b) integration variables in the disk plane.}
\end{figure}

The chosen $\sigma(r)$ is continuous since the gravitational potential and the rotational speed would diverge at points of discontinuity. That is directly related to the assumed zero thickness of the disk. The form of $\sigma(r)$ does not have a physical significance. Its solely purpose is to test the ability of the method to reconstruct it from an incomplete velocity profile. The constants in $\sigma(r)$ are chosen so that the radial derivative of the produced potential is always positive i.e. the gravitational force is towards the center not away from it. The piece for $0<r\le 2$ will test the method's ability to reconstruct small details in $\sigma(r)$, the decreasing piece for $4<r\le 6$ is a form that produces a flat velocity curve by itself, the piece for $4<r\le 6$ will test how the method reconstructs a planar halo, and the piece for $6<r\le 8$ is just a continuous drop to zero. 


The total mass is
\begin{equation}
M =  \int_0^\infty  \sigma(r)2\pi r dr = 308.6.
\end{equation}
The quantity relevant to the amount of mass within certain radii is $r \sigma(r)$ not $\sigma(r)$ alone. That amplifies the mass density at bigger radii and the $4<r\le 8$ part of the mass density actually contains 40\% of the total mass although it looks negligible on the $\sigma(r)$ plot.

Before testing the method, it is necessary to compute the potential and the real velocity curve from the given toy mass density. It is possible to use the derived Bessel spectrum formulas in reverse mode but we chose an independent method - direct numerical integration. The gravitational potential at radius $r$ on the disk is (see Fig.1(b))
\begin{equation}
\Phi(r,z=0)=\int \frac{-G dm}{\rho} = -G \int_{\phi=0}^{2\pi} \int_{R=0}^{\infty} \frac{\sigma(R) \;R\; dR\; d\phi}{\sqrt{R^2+r^2-2 R r cos\phi}}.
\end{equation}
This form is problematic for numerical integration at the points where the denominator of the integrand goes to zero. That zero is canceled by the numerator but most numerical routines are likely to have severe cancellation errors.  The problem is eliminated if $dm$ is expressed in terms of coordinates, $(\rho, \theta)$, with respect to the point where the potential is being computed (see Fig.1(b)) since that cancels out the denominator:
\begin{equation}
\Phi(r,z=0)=\int \frac{-G dm}{\rho} = -G \int_{\theta=0}^{2\pi} \int_{\rho=0}^{\infty} \frac{\sigma(R) \rho d\rho d\theta}{\rho}= -G \int_{\theta=0}^{2\pi} \int_{\rho=0}^{\infty} \sigma(R) \,d\rho \,d\theta,
\end{equation}
where $R = \sqrt{r^2+\rho^2 - 2\, r\, \rho \, cos\theta}$.

\begin{figure}[h]
\includegraphics[width=7.1cm]{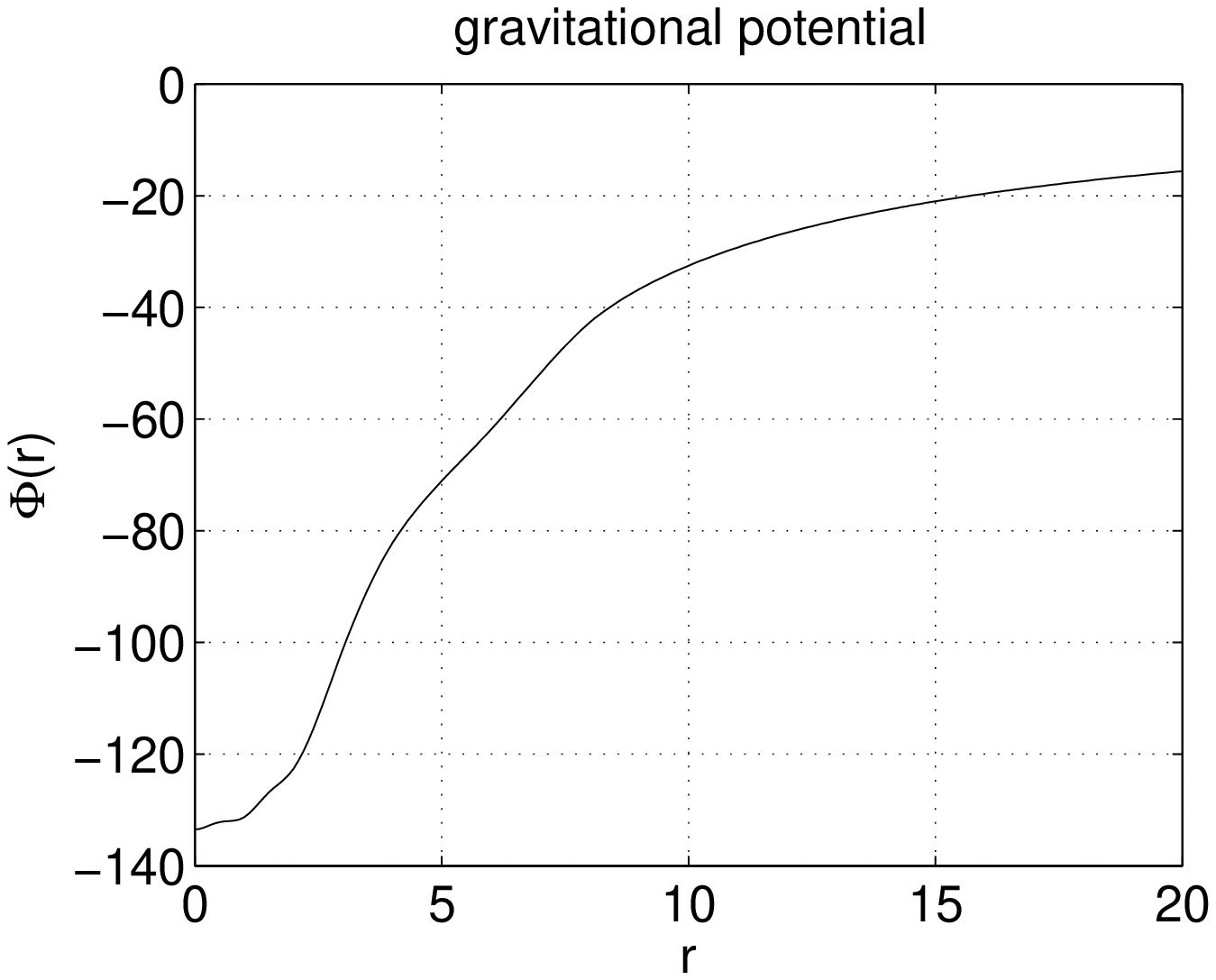}
\includegraphics[width=7.1cm]{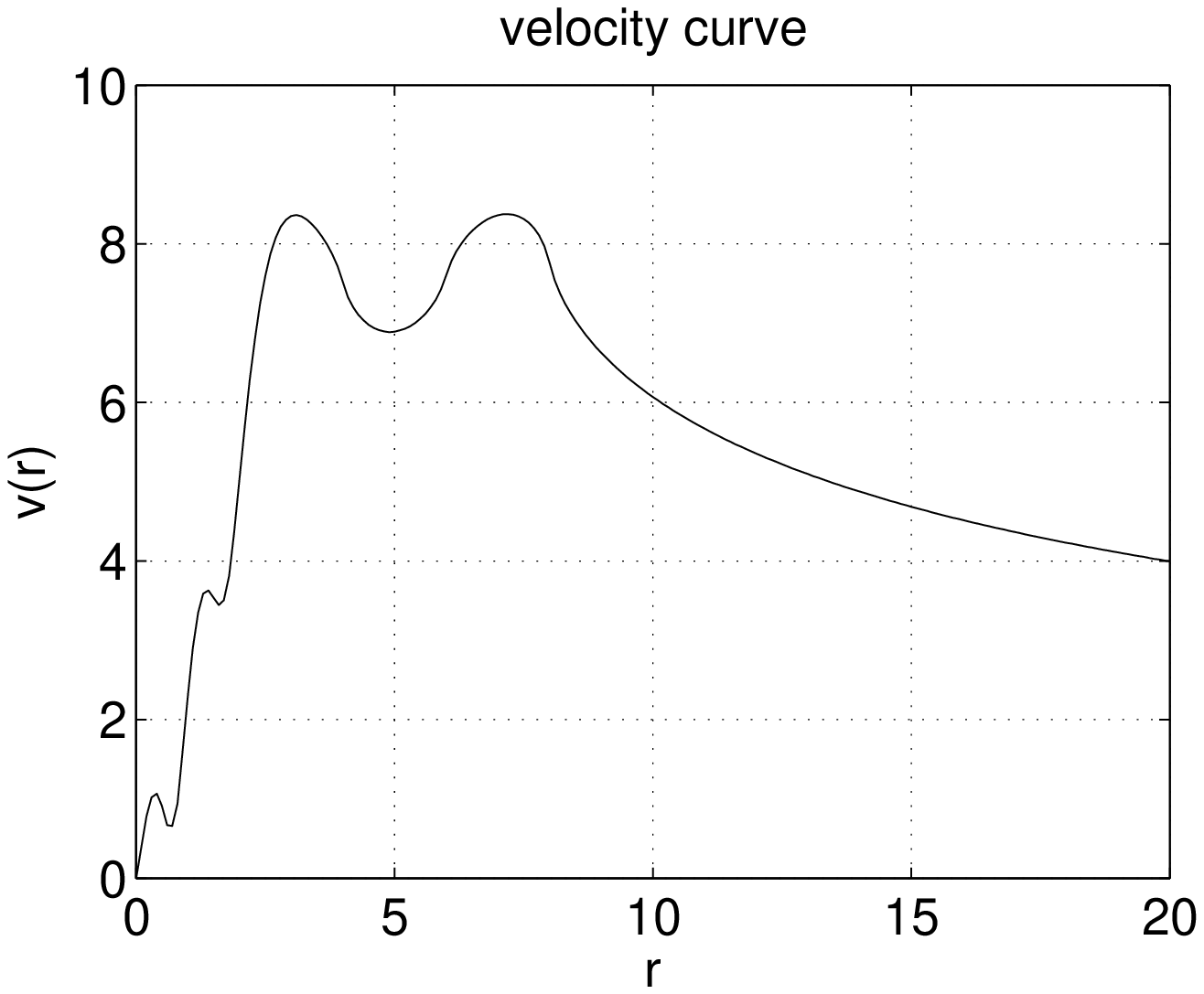}
\caption{toy mass density: (a) gravitational potential 
(b) velocity curve. }
\end{figure}

Using equation (19), the potential was calculated numerically  for radii from 0 to 20 with a step of $\Delta r = 0.1$ and interpolated with splines (shown on Fig 2(a)). The derivative of the spline-fitted potential was computed numerically and then the velocity curve (shown on Fig.2(b)) was calculated according to
\begin{equation}
v(r) = \sqrt{r\,\frac{\partial\Phi(r,z=0)}{\partial r}}.
\end{equation} 

In the following, we test how well the method will reconstruct the toy mass density if part of the velocity curve is not known, hence extrapolated.

In the first test we assume that the velocity curve is known out to $r_{max}=10$. The total mass estimate obtained from equation (12) is $M_e = 368.2$ which overshoots the real mass $M=308.6$ The velocity curve was extrapolated after  $r_{max}$ (see Fig.3(a)) by formula (13) and fed to the calculation procedure outlined in the previous section. The resulting reconstructed mass density is shown on Fig.3(b). It is so close to the original one that they are hard to distinguish. The original and the reconstructed integrated masses for $r<r_{max}$ are, respectively, 308.6 and 319.1. The higher reconstructed mass is caused by the overshooting of the total mass estimate $M_e$.

\begin{figure}[h]
\includegraphics[width=7.1cm]{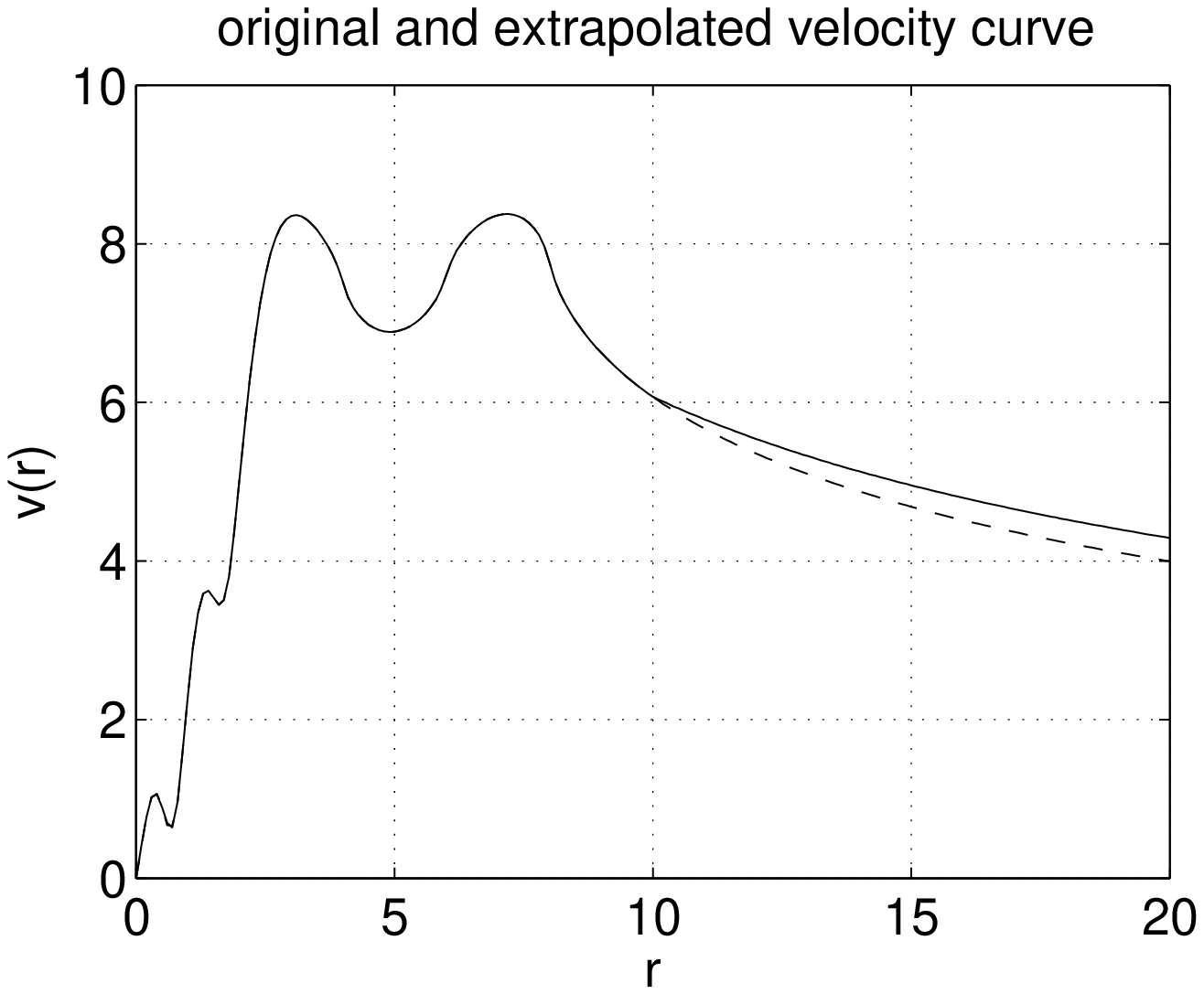}
\includegraphics[width=7.1cm]{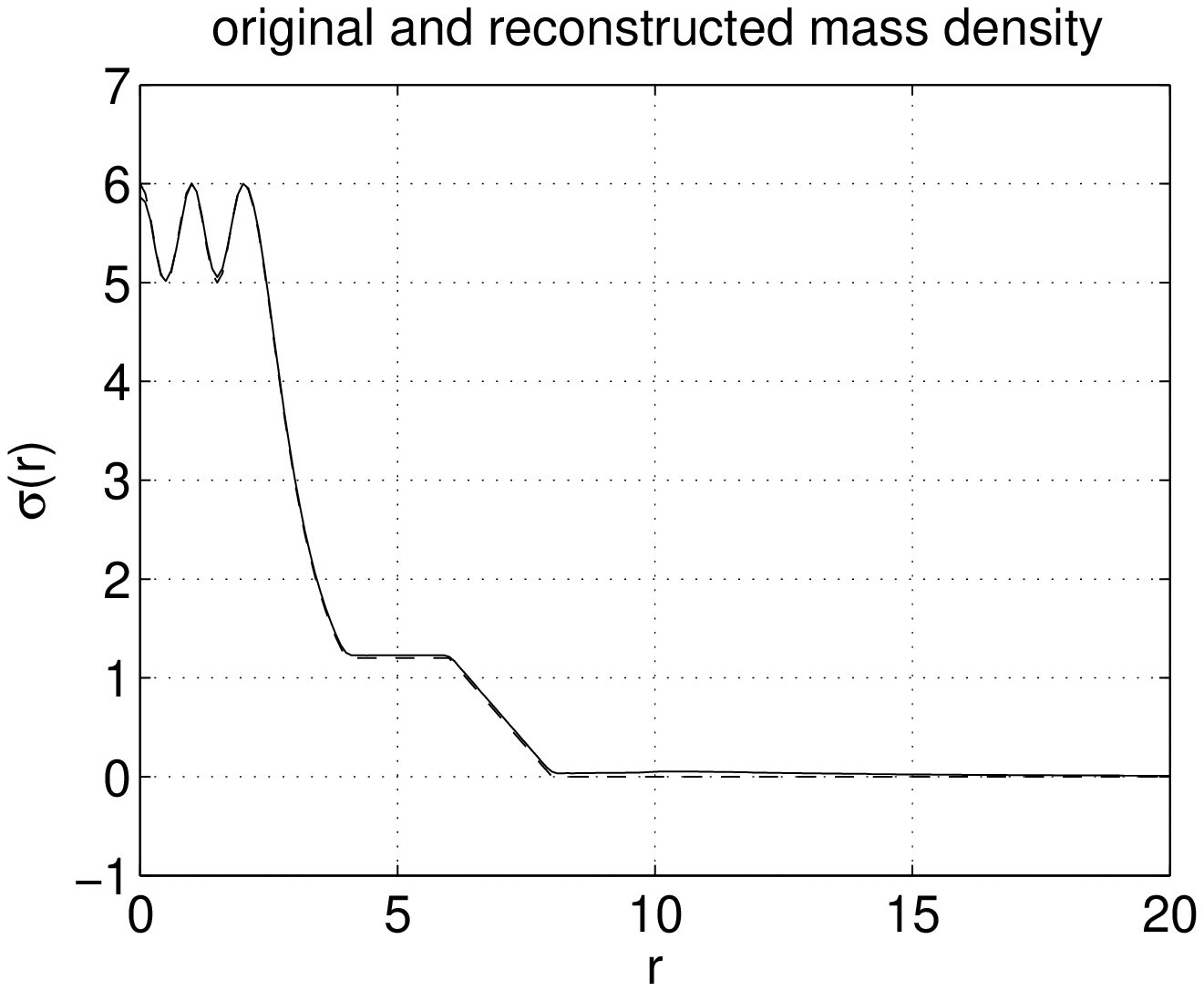}
\caption{(a) original (dashed) and extrapolated (solid) velocity curve for $r_{max}=10$
(b) original (dashed) and reconstructed (solid) mass density. }
\end{figure}

In the second test we assume that the velocity curve is known out to $r_{max}=4$. That leaves 40\% of the original mass outside $r_{max}$. The total mass estimate is $M_e = 226.0$ which is below the real mass $M=308.6$. The extrapolated velocity curve (see Fig.4(a)) has lost one of the 'humps' and is below the original one for $r>r_{max}$. Due to that we expect a reconstructed mass density below the original one and a poor reconstruction of the mass density shape for $r>r_{max}$. The reconstructed mass density shown on Fig.4(b) indeed preserves the shape of the original one although it is slightly below it for $r<r_{max}$. The reconstructed mass inside $r_{max}$ is 158.5 while the original mass inside $r_{max}$ is 182.9. The discrepancy again is explained by the low mass estimate $M_e$.

\begin{figure}[h]
\includegraphics[width=7.1cm]{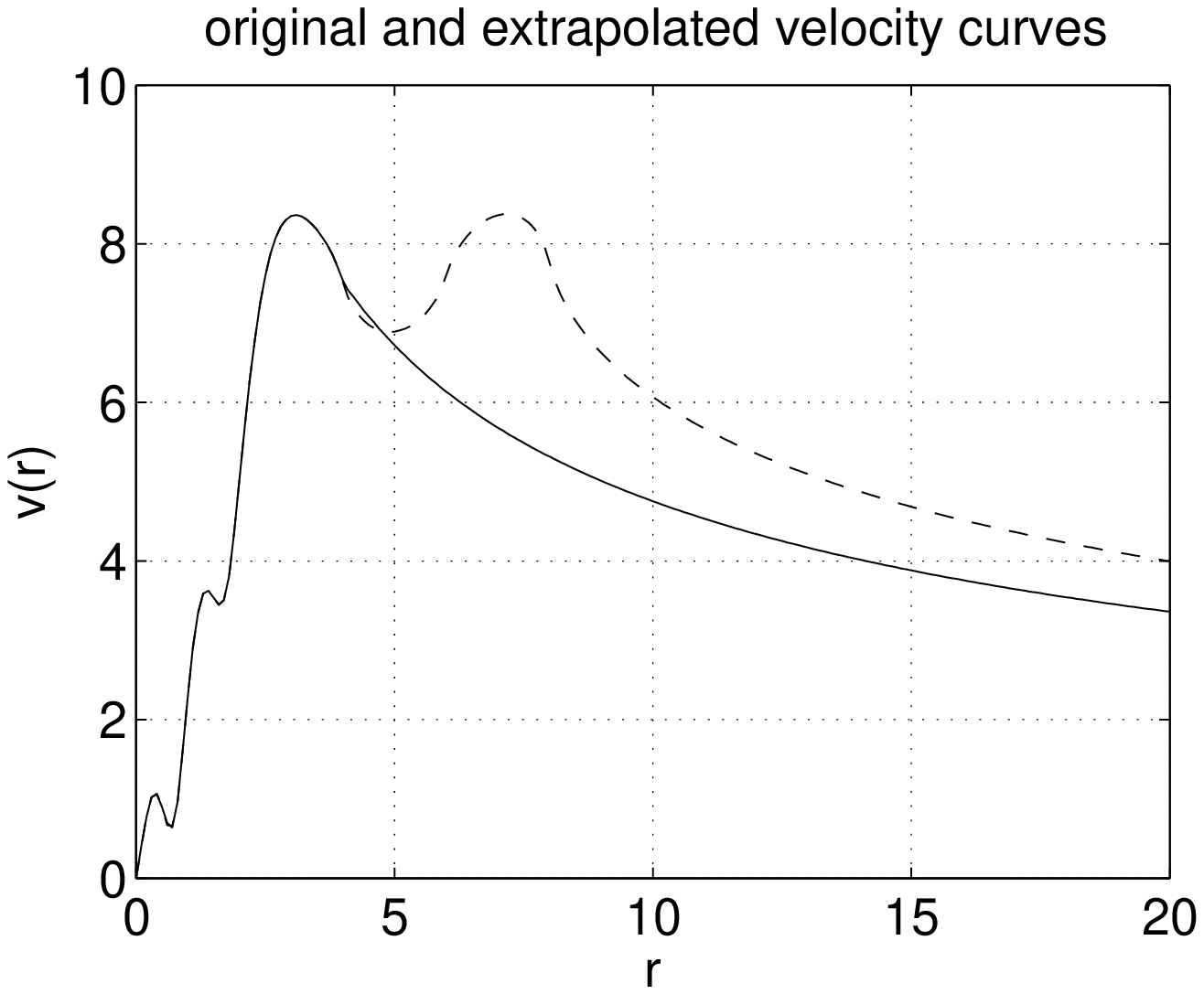}
\includegraphics[width=7.1cm]{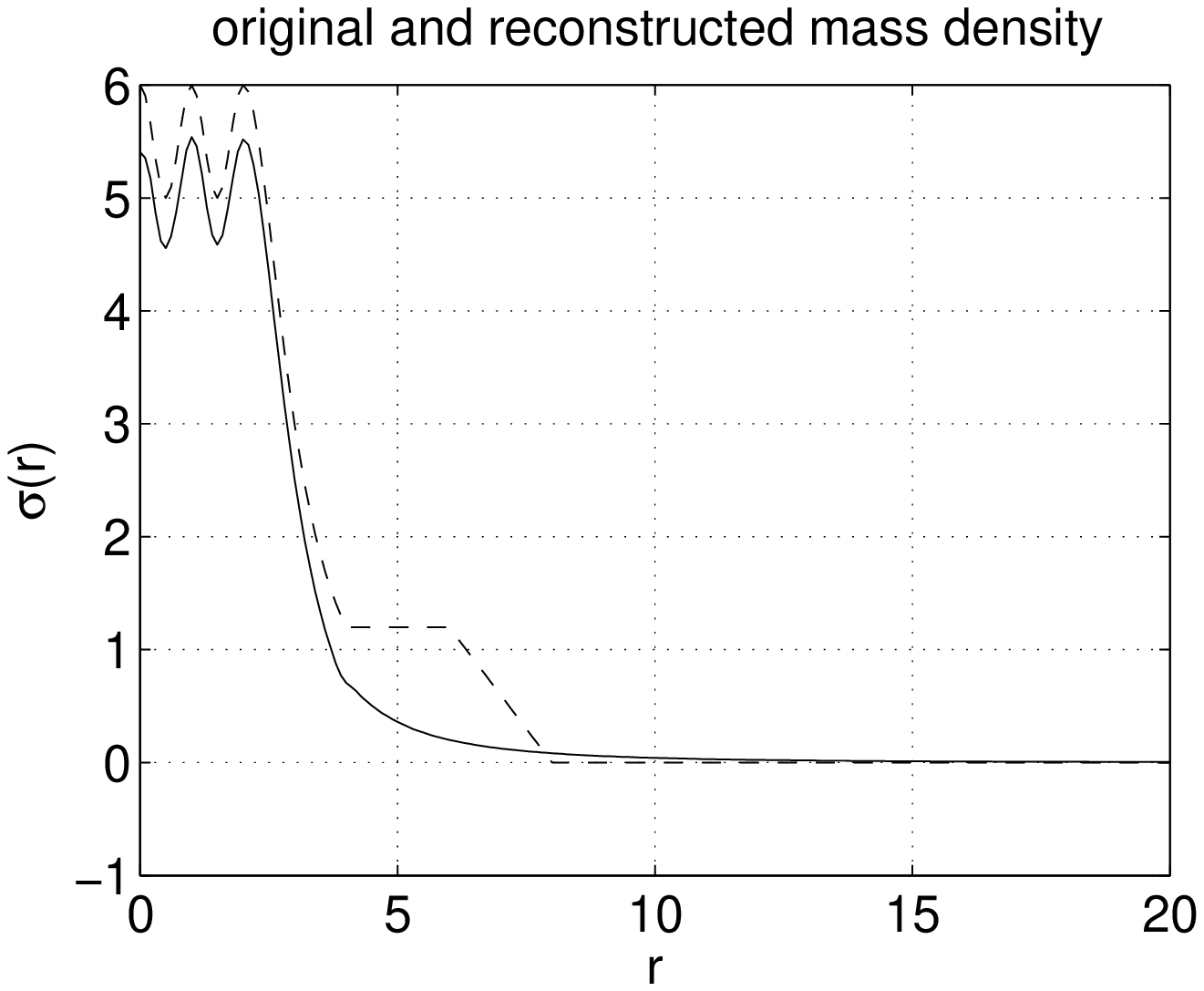}
\caption{(a) original (dashed) and extrapolated (solid) velocity curve for $r_{max}=4$
(b) original (dashed) and reconstructed (solid) mass density. }
\end{figure}

In the third test we assume that the velocity curve is known out to $r_{max}=2$. That leaves 61\% of the original mass outside $r_{max}$. The total mass estimate is $M_e = 50.6$ which is far below the real mass $M=308.6$. The extrapolated velocity curve (see Fig.5(a)) has lost a lot of details and height for $r>r_{max}$. Due to that we expect the reconstructed mass density to be severely below the original one and poor reconstruction of the mass density shape for $r>r_{max}$. The reconstructed mass density shown on Fig.5(b) indeed preserves somewhat the shape of the original one although is very much below it for $r<r_{max}$. The shape is totaly washed out for $r>r_{max}$. The reconstructed mass inside $r_{max}$ is 27.4 while the original mass inside $r_{max}$ is 69.1. 

\begin{figure}[h]
\includegraphics[width=7.1cm]{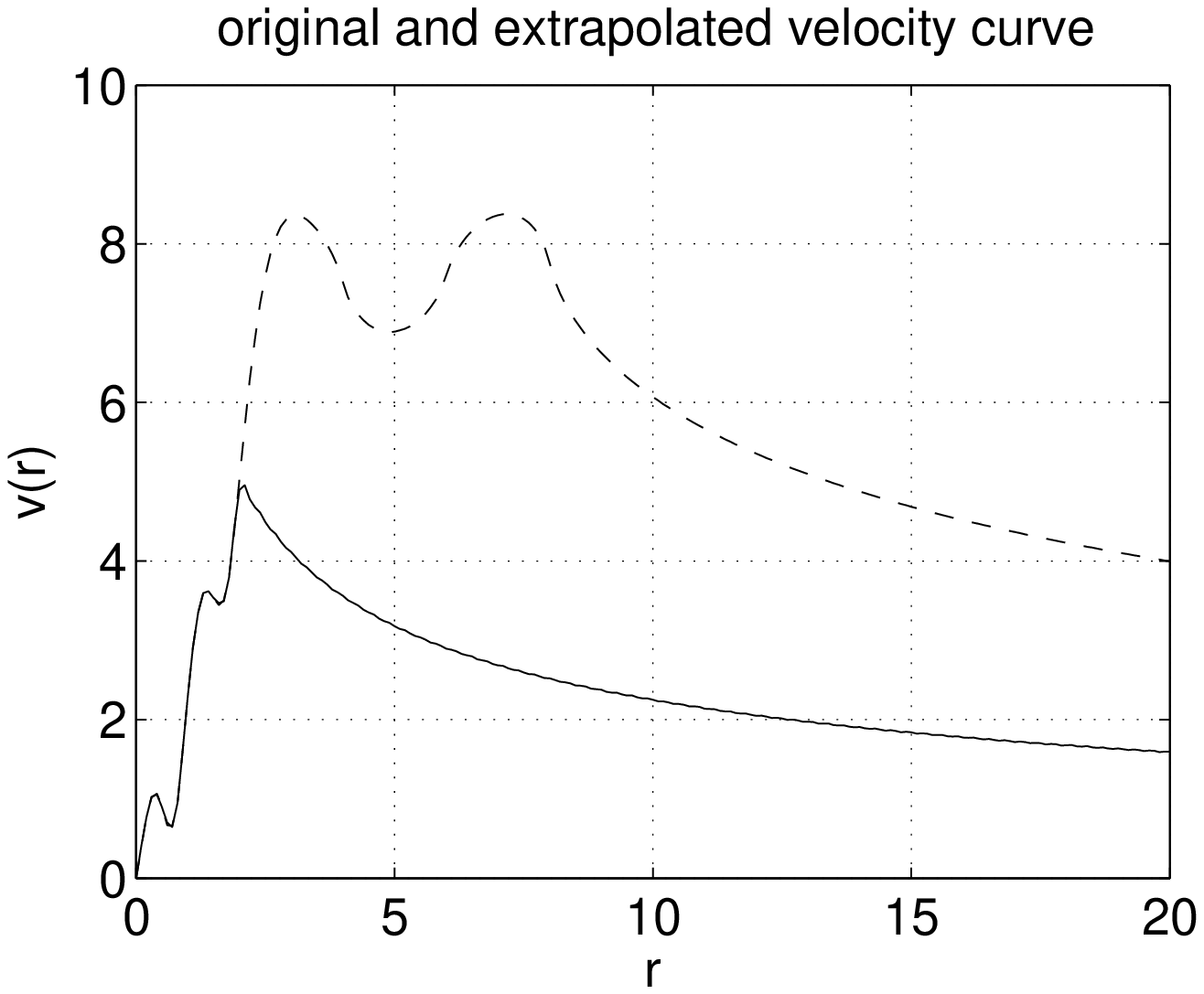}
\includegraphics[width=7.1cm]{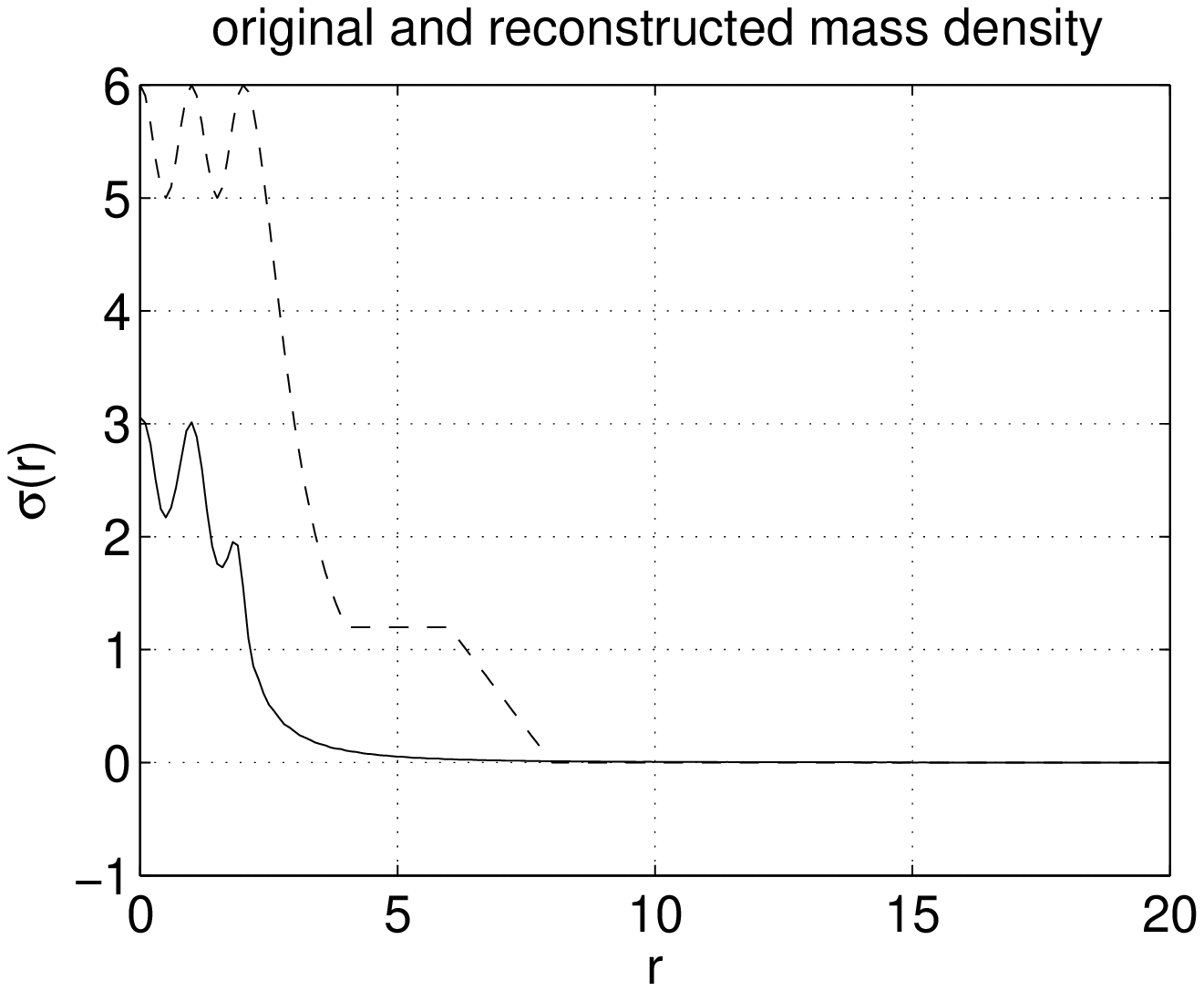}
\caption{(a) original (dashed) and extrapolated (solid) velocity curve for $r_{max}=2$
(b) original (dashed) and reconstructed (solid) mass density. }
\end{figure}

The tests show that the total mass estimate $M_e$ is close to the real mass when $r_{max}$ is on the Keplerian tail of the velocity curve (test 1) or at least near it (test 2) i.e. when a big part of the total mass is inside $r_{max}$. The height of the reconstructed mass density depends on $M_e$ but is not extremely sensitive to it - $M_e$ is off by 20\% in test 1, yet the reconstructed distribution is indistinguishable from the original. The shape of the reconstructed mass density inside $r_{max}$ is fairly independent of the goodness of the mass estimate. The shape outside $r_{max}$ is washed out due to the velocity curve extrapolation (tests 2 and 3).

\section{APPLICATION TO SPIRAL GALAXIES}
The velocity curve fits of several galaxies are taken from \citet{cop}.

The velocity curve of Milky Way is known from HI observations out to $r_{max}=30 \; kpc$.  The total mass estimate of equation (12) is proportional to the extent of the flat part of the velocity curve, which is currently unknown. We can speculate that the Keplerian tail starts right after $30 \; kpc$, which gives a total mass estimate of $M_e = 30 \times 10^{10} \; M_\odot$. The corresponding extrapolated velocity curve and the reconstructed mass density are shown on Fig. 6. The visible disk has a radius of $15 \; kpc$. The reconstructed mass  is $15 \times 10^{10} \; M_\odot$ (half of the total mass) inside the visible disk and $23 \times 10^{10} \; M_\odot$ inside $r_{max}$. This is in contrast to the widespread view that the mass must be concentrated predominantly outside the visible disk. The conventional value of the Milky Way's absolute magnitude is -20.5 which gives a luminosity of $L_V = 1.36 \times 10^{10} L_{V\odot}$. The mass-to-light ratio is $M/L_V = 11 \; M_\odot/L_{V\odot}$ for the visible disk and $22 \; M_\odot/L_{V\odot}$ for the whole galaxy. 

\begin{figure}[h]
\includegraphics[width=7.1cm]{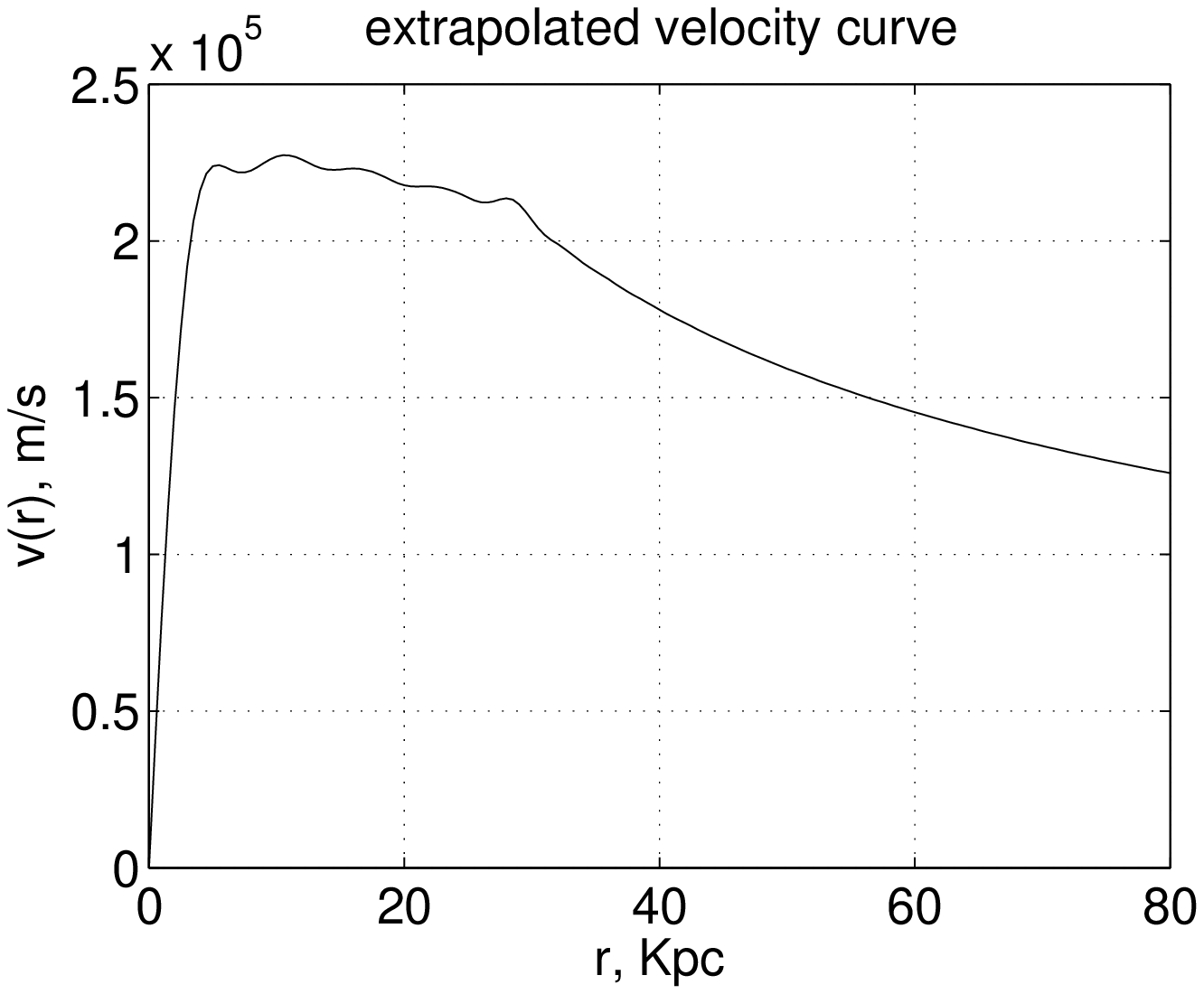}
\includegraphics[width=7.1cm]{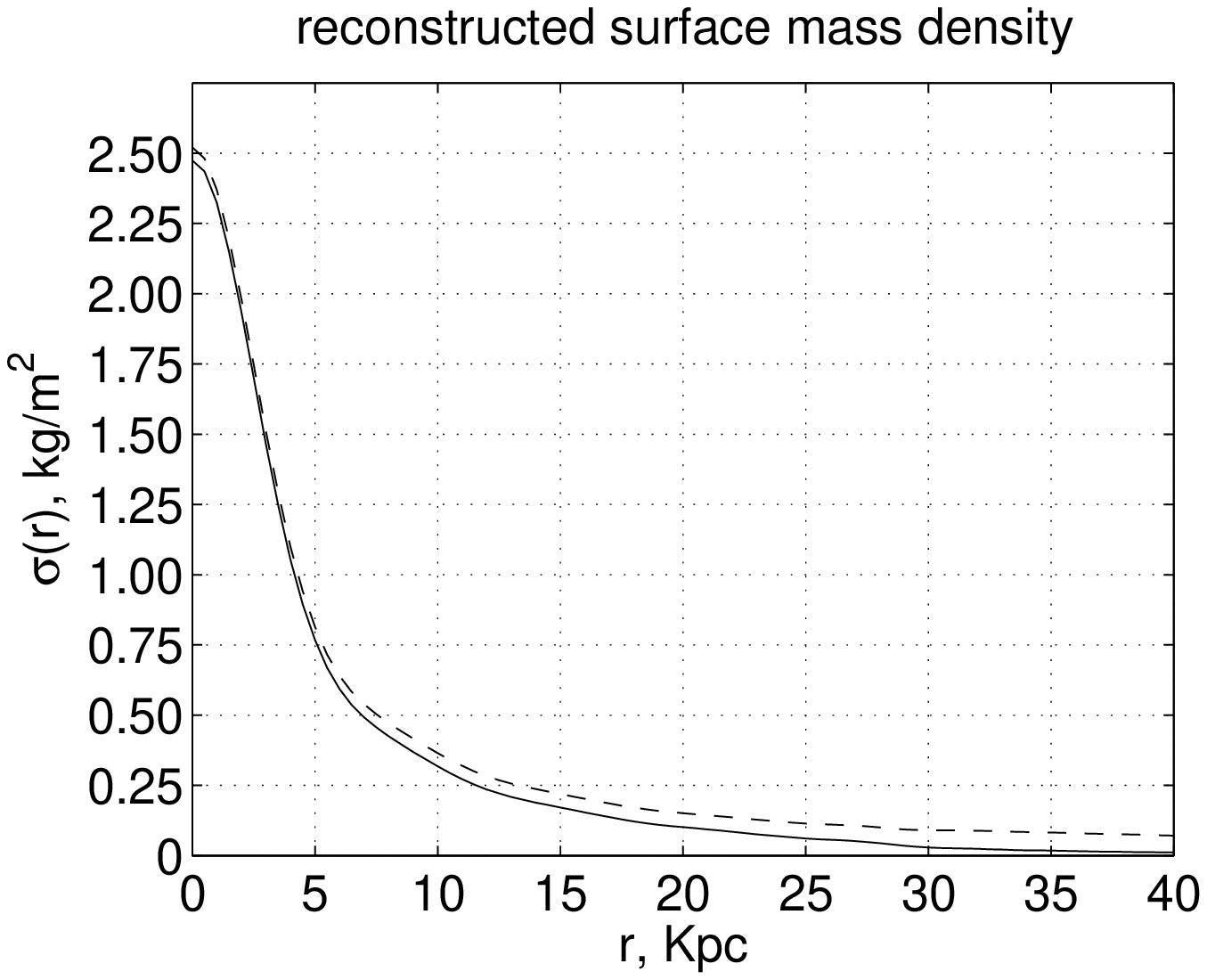}
\caption{Milky Way: (a) velocity curve extrapolated after $r_{max}=30 \; kpc$
(b) reconstructed mass density for $r_{max}=30 \; kpc$ (solid) and $r_{max}=200 \; kpc$ (dashed). }
\end{figure}

To evaluate the effect of the unknown extent of the flat part of the velocity curve, we continued the experimental curve with a flat line till $r_{max}=200 \;kpc$ and extrapolated with a Keplerian tail after that. That gives a total mass estimate $M_e = 197 \times 10^{10} \; M_\odot$. The resulting reconstructed mass density is the dashed curve on Fig. 6(b). The reconstructed mass inside the visible disk is $16 \times 10^{10} \; M_\odot$ and inside $30 \;kpc$ is $23 \times 10^{10} \; M_\odot$. It is remarkable, that although the total mass estimate increased more than 6 times, the masses and the mass density inside the experimentally known part of the velocity curve ($r<30 \;kpc$) remained fairly unchanged.

Another well studied and measured galaxy is NGC 3198 whose velocity curve is known out to $r_{max}=30 \; kpc$. Assuming the Keplerian decay starts right after $r_{max}$, gives $M_e = 15 \times 10^{10} \; M_\odot$. The extrapolated velocity profile and the resulting mass density are shown on Fig. 7. The visible disk has a radius of $14 \; kpc$. The reconstructed mass inside the visible disk is $6.5 \times 10^{10} \; M_\odot$ (43\% of the total mass) and $11 \times 10^{10} \; M_\odot$ inside $r_{max}$ . The V-band luminosity of NGC 3198 is $L_V = 0.7 \times 10^{10} L_{V\odot}$ \citep{3198}. The mass-to-light ratio is $M/L_V = 9.3 \; M_\odot/L_{V\odot}$ for the visible disk and $21 \; M_\odot/L_{V\odot}$ for the whole galaxy. 

Continuing the flat part of the velocity curve with a straight line till $r_{max}=200 \;kpc$ and extrapolating with a Keplerian tail after that gives $M_e = 100 \times 10^{10} \; M_\odot$. The resulting mass density is the dashed line on Fig. 7(b). The corresponding mass inside the visible disk is $7.2 \times 10^{10} \; M_\odot$ and $15 \times 10^{10} \; M_\odot$ inside $30 \;kpc$. Again, the mass distribution inside $r_{max}=30 \; kpc$ is relatively insensitive to the extent of the velocity curve flat part.

\begin{figure}[h]
\includegraphics[width=7.1cm]{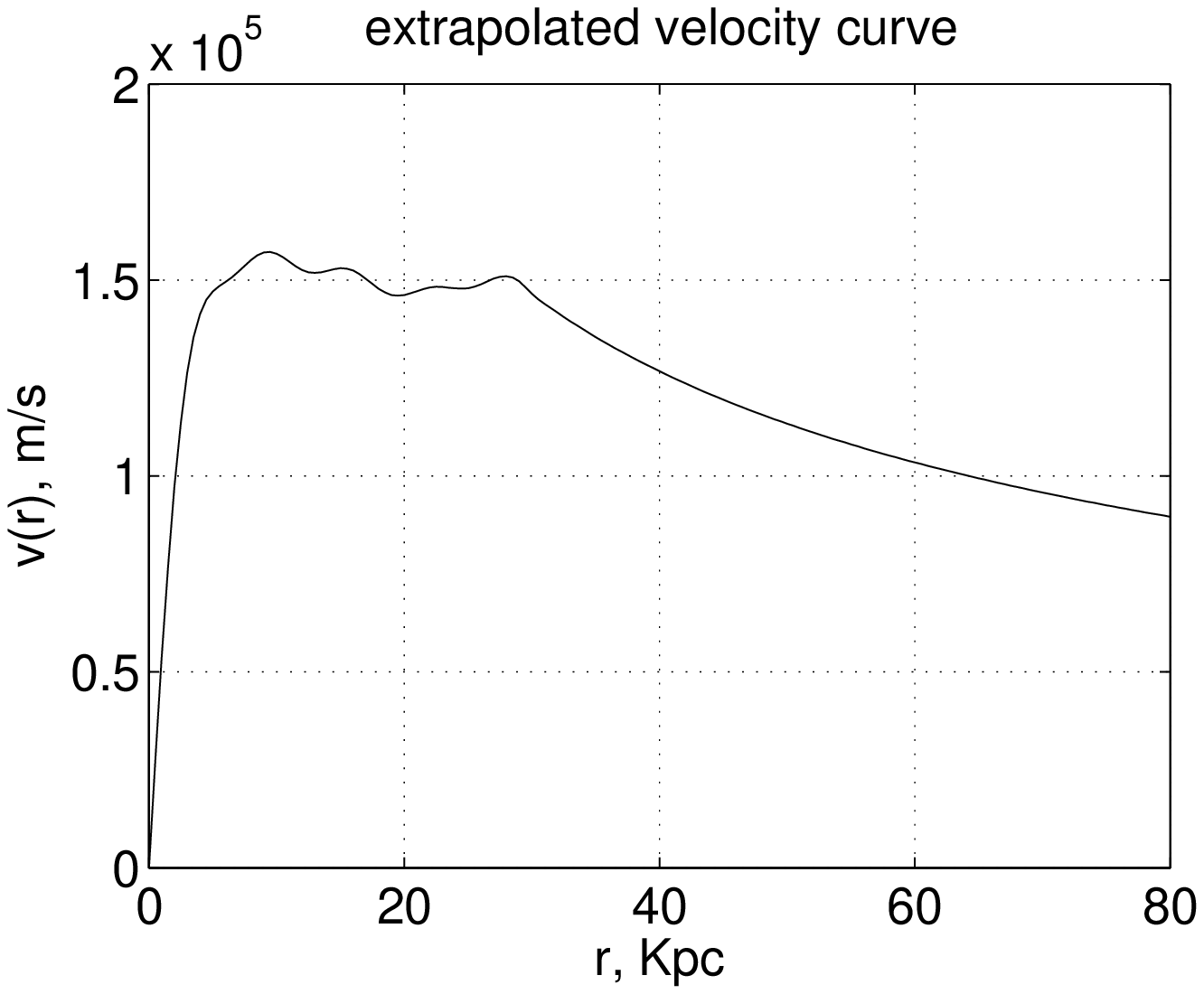}
\includegraphics[width=7.1cm]{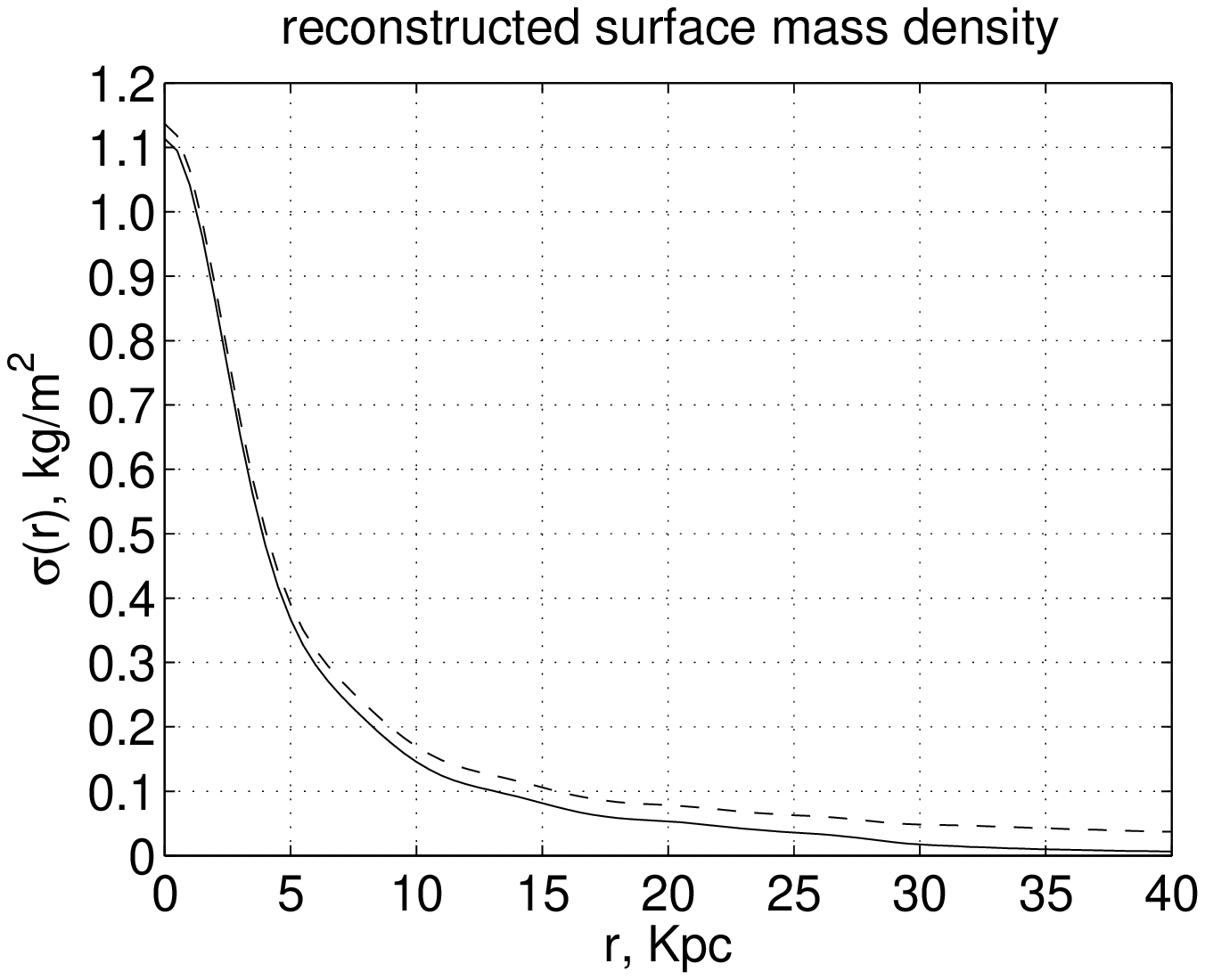}
\caption{galaxy NGC 3198: (a) velocity curve extrapolated after $r_{max}=30 \; kpc$
(b) reconstructed mass density for $r_{max}=30 \; kpc$ (solid) and $r_{max}=200 \; kpc$ (dashed). }
\end{figure}

Last, we consider the galaxy NGC 3031 whose velocity curve is known out to $r_{max}=21 \; kpc$. It is an excellent candidate for extrapolation since it seems to show  a Keplerian decline already. The extrapolated part of the NGC 3031 velocity curve continues very naturally the slope of the experimental curve (see Fig. 8(a)). We can expect an excellent reconstruction of the mass density curve possibly with some features outside $r_{max}$ washed out. The resulting surface mass density is shown on Fig.8(b). The total mass estimate is $M_e=14 \times 10^{10} \; M_\odot$. The reconstructed mass inside $r_{max}$ is $11 \times 10^{10} \; M_\odot$ (79\% of the total mass).

\begin{figure}[h]
\includegraphics[width=7.1cm]{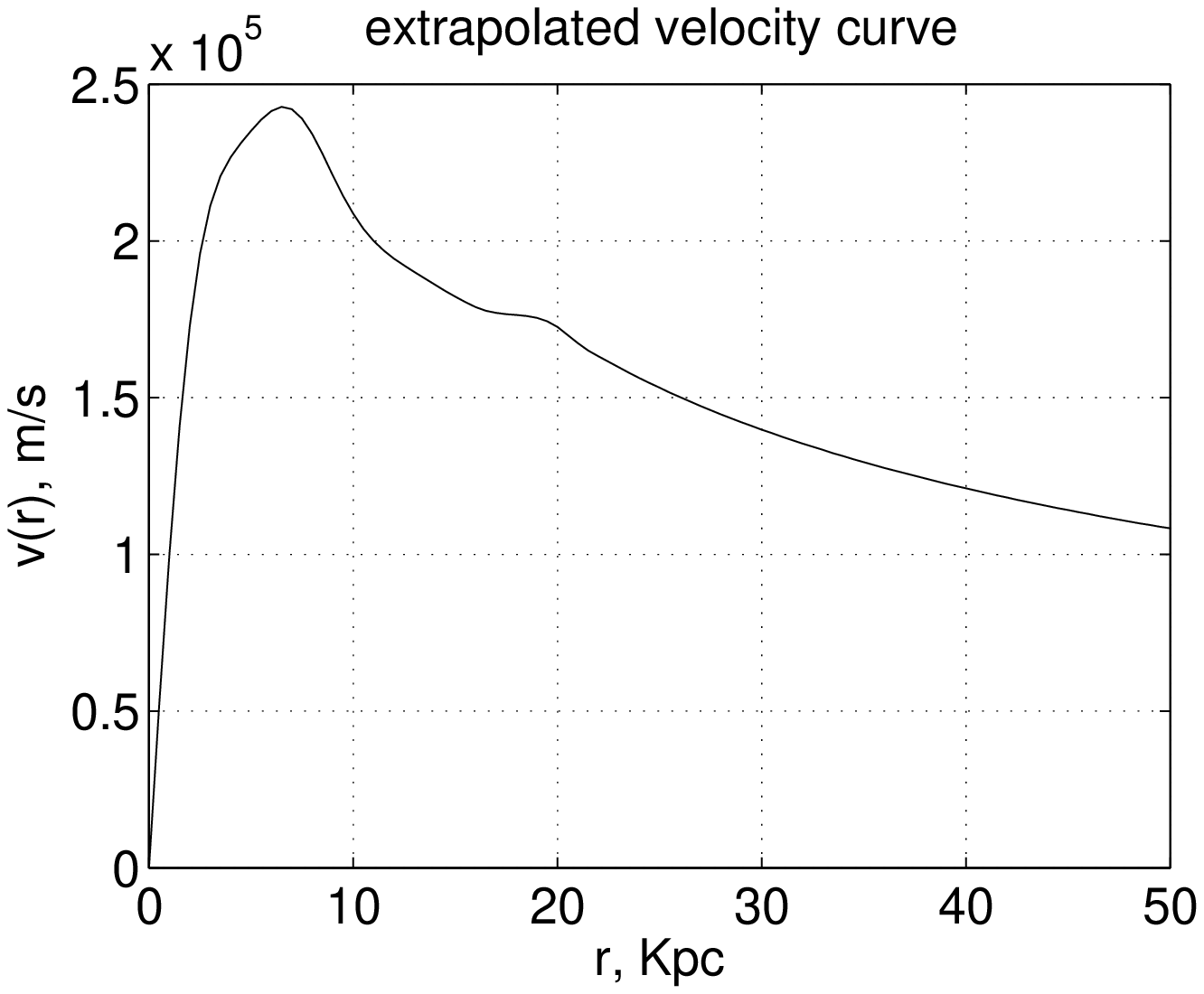}
\includegraphics[width=7.1cm]{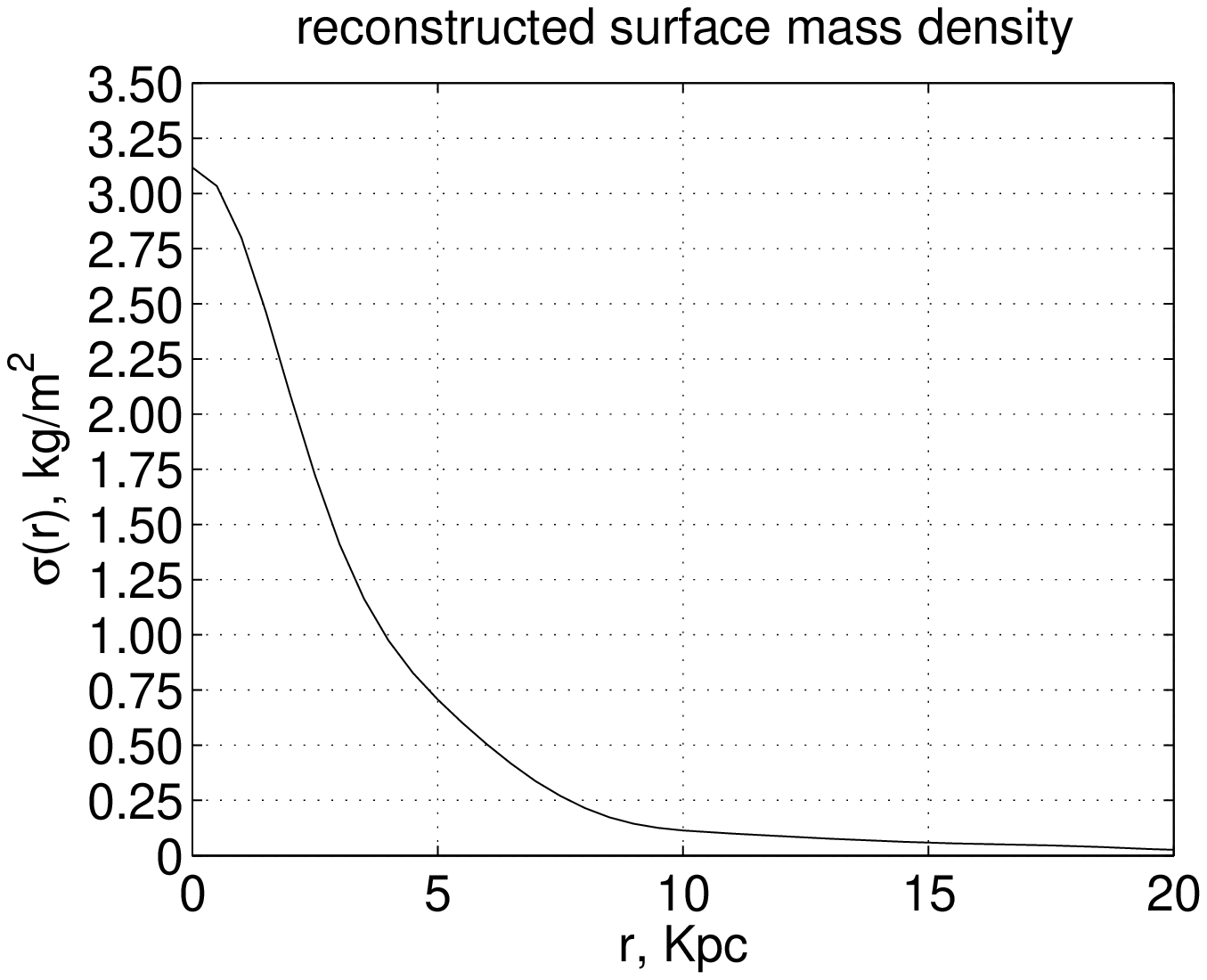}
\caption{galaxy NGC 3031: (a) velocity curve extrapolated after $r_{max}=21 \; kpc$
(b) reconstructed mass density. }
\end{figure}

\section{DISCUSSION}

For a spherically symmetric mass distribution, the mass density inside some radius can be calculated without knowing the velocity curve outside that radius. In the axial symmetry of the disk model, that is generally not true. Nevertheless, the reconstructed mass densities inside $r=30\;kpc$ were shown in the previous section to be rather insensitive to the extent of the flat part of the velocity curve which was varied from $r_{max}=30 \; kpc$ to $200 \; kpc$. We can regard the obtained mass densities and the total masses inside $r_{max}=30 \; kpc$ as fairly good estimates of the real ones unless the model is not applicable or the velocity curves start to increase after $r_{max}$.

The high mass-to-light ratios obtained, even when the velocity curves were assumed to decay after the current experimental $r_{max}$, are suggestive of the presence of dark matter. What fraction of the matter is dark or nonbarionic remains to be seen.

The estimates of the total galactic masses are roughly proportional to the extent of the flat parts of the velocity curves, which are currently unknown. While the mass inside the visible disk remains relatively constant, the mass outside it increases with increasing the total galactic mass.

If applicable, the flat disk model with velocity extrapolation, is expected to give fairly accurate estimates both of the total mass and the surface mass density function for galaxies with already decaying velocity curves like NGC 3031.

\section{ACKNOWLEDGEMENTS}
I am grateful to Martin Tzanov for his helpful criticism of the original manuscript.


\begin{thebibliography}{}
\bibitem[van Albada et al.(1985)]{3198} van Albada, T.S., Bahcall, J.N., Begeman, K. and Sanscisi, R., 1985. Ap. J. 295, 305
\bibitem[Cooperstock and Tieu(2005)]{cop} Cooperstock, F. I., \& Tieu, S. 2005, preprint (astro-ph/0507619)
\bibitem[Binney and Merrifield(1998)] {textbook} Binney, J. \& Merrifield, M. 1998, Galactic Astronomy, 
\bibitem[Jackson(1998)]{jak} Jackson, D. 1998, Classical Electrodynamics (3rd ed.)
\bibitem[Kent(1987)] {kent2} Kent, S. 1987, Astrophys. J., 93, 816
\bibitem[Milgrom(1983)] {mond} Milgrom, M. 1983, Ap. J., 270, 365
\bibitem[Nordsieck(1973)] {nord} Nordsieck, K. 1973, Ap. J., 184, 719
\bibitem[Takamiya and Sofue(2000)]{soph} Takamiya T. \& Sofue Y. 2000, Ap. J., 534, 670
\bibitem[Toomre(1963)] {toom} Toomre, A. 1963, Ap. J., 138, 385
\end{thebibliography}
\end{document}